\documentclass{article}

\usepackage{PRIMEarxiv}

\usepackage[utf8]{inputenc} % allow utf-8 input
\usepackage[T1]{fontenc}    % use 8-bit T1 fonts
\usepackage{amsmath}        % advanced math typesetting
\usepackage{amssymb}        % extended math symbols (includes amsfonts)
\usepackage{amsfonts}       % blackboard bold, etc. (often loaded by amssymb, but kept for clarity)
\usepackage{mathtools}      % extra math tools
\usepackage{amsthm}         % theorem environments
\usepackage{bm}             % bold math symbols

\usepackage{graphicx}       % graphics support
\usepackage{float}          % improved figure placement
\usepackage{caption}        % custom captions
\usepackage{subcaption}     % subfigures and subtables
\usepackage{tikz}           % TikZ figures
\usetikzlibrary{arrows.meta,positioning,shapes.geometric,calc,fit,backgrounds}

\usepackage{booktabs}       % professional-quality tables
\usepackage{tabularx}       % flexible-width tables
\usepackage{multirow}       % multirow cells in tables
\usepackage{adjustbox}      % fine-tuning table/figure alignment

\usepackage{algorithm}
\usepackage{algorithmic}

\usepackage{listings}

\usepackage{microtype}      % microtypography (improves spacing)
\usepackage{textcomp}       % additional text symbols
\usepackage{nicefrac}       % compact fractions like 1/2
\usepackage{siunitx}        % typesetting SI units and numbers

\usepackage{url}            % simple URL formatting
\usepackage{hyperref}       % hyperlinks (should be loaded last, with few exceptions)
\usepackage{xcolor}         % color support (required by hyperref if custom link colors are used)
\usepackage[capitalize,noabbrev]{cleveref} % must be loaded after hyperref

\usepackage{geometry}       % page layout
\usepackage{fancyhdr}       % custom headers and footers

\usepackage{physics}        % For macros like norm

\usepackage{cite}           % improved citation handling
\usepackage{lipsum}         % dummy text (remove in final document)
\usepackage{pdfpages}       % include PDF pages
\usepackage{orcidlink}      % ORCID icons with hyperlinks

\usepackage{array}
\newcolumntype{M}[1]{>{\centering\arraybackslash}m{#1}}

\theoremstyle{plain}

\theoremstyle{definition}

\theoremstyle{remark}

\definecolor{pBlue}{HTML}{D6E6F5}   \definecolor{pBlueE}{HTML}{6F9AC6}
\definecolor{pLav}{HTML}{E6DFF4}    \definecolor{pLavE}{HTML}{8E7DBE}
\definecolor{pYel}{HTML}{FFF1C1}    \definecolor{pYelE}{HTML}{C9A94A}
\definecolor{pPeach}{HTML}{FDE0CC}  \definecolor{pPeachE}{HTML}{D08B5E}
\definecolor{pGreen}{HTML}{D8EFD9}  \definecolor{pGreenE}{HTML}{6FAE72}
\definecolor{pPink}{HTML}{F8D7E1}   \definecolor{pPinkE}{HTML}{C9748F}
\definecolor{pGray}{HTML}{E9ECEF}   \definecolor{pGrayE}{HTML}{8A939C}
\definecolor{pInk}{HTML}{3F4A5A}

\title{Demographic Encoding is Not Exposure: Fairness and Privacy in Chest X-ray Foundation Models}

\makeatletter
\newcommand{\affilmark}[1]{\textsuperscript{#1}}
\makeatother

\author{
  \normalfont \textbf{Sahar Askari}\,\text{\orcidlink{0009-0006-3933-5432}}\affilmark{a}
  \and
  \normalfont \textbf{Amin Golnari}\,\text{\orcidlink{0000-0002-3841-0588}}\affilmark{b}
}

\begin{document}
\maketitle

\footnotetext[1]{%
  Corresponding author. Email: \texttt{askarisahar92@gmail.com}. \\
  Department of Computer Science and Engineering, Shiraz University, Shiraz, Iran.
}

\footnotetext[2]{%
  Email: \texttt{amin.g@vinuni.edu.vn}. \\
  College of Engineering \& Computer Science, VinUniversity, Hanoi, Vietnam.
}

\begin{abstract}
  Medical foundation models encode patient demographics and can reveal whether a patient was in the training set, and these properties are often treated as one problem. We tested whether they are aligned by measuring demographic encoding, diagnostic fairness gaps and membership-inference vulnerability on the same models. We trained 200 linear or multilayer-perceptron (MLP) heads on frozen BiomedCLIP, DINOv2 and RAD-DINO embeddings, with three seeds, and audited them on CheXpert and NIH ChestX-ray14. Demographic encoding explained almost no variance in patient vulnerability, and all 72 CheXpert associations were statistically equivalent to zero. Erasing linearly decodable demographics left leakage unchanged but reduced age-related underdiagnosis gaps. Leakage depended instead on head capacity and training length. On RAD-DINO, a linear head reached a membership-inference area under the curve (AUC) of 0.694, compared with 0.907 for a 30-epoch MLP, at similar test performance. Fairness and privacy should therefore be audited separately, and linear heads or shorter training reduce leakage with little or no loss of utility. The code is available on \href{https://github.com/SAMZ2610/privacy_bias_fairness}{GitHub} for public use.
\end{abstract}

% keywords

%=====================================================================
\section{Introduction}\label{sec:intro}
Foundation models are becoming the default starting point for medical imaging, and they raise two concerns for clinical use. The first concern is fairness. Chest X-ray classifiers underdiagnose under-served patient groups~\cite{seyyedkalantari2021,seyyedkalantari2021chexclusion}. Image representations also encode protected attributes such as sex, age and self-reported race, and race in particular cannot be read from the image by clinicians~\cite{gichoya2022,glocker2023}. The second concern is privacy. Membership inference attacks (MIAs) can reveal whether a patient's record was used for training~\cite{shokri2017,carlini2022}. A recent large-scale audit showed that this risk is unevenly distributed across patients and grows with model capacity~\cite{knolle2026}.

These two concerns are usually discussed together, with an implicit causal story. If a representation encodes who the patient is, that patient may be easier to single out, and the encoded attributes may also provide a shortcut for biased predictions~\cite{geirhos2020}. This story predicts that demographic encoding, diagnostic fairness gaps and membership vulnerability are aligned. The question has practical consequences, because it determines whether one mitigation, such as removing demographic information, serves both goals. If the three quantities are aligned, one mitigation could serve both goals. If they are not, each risk needs its own audit. Related studies have examined encoding versus its linear use for prediction~\cite{kim2026}, persistent demographic information in X-ray embeddings~\cite{santos2025}, fairness-privacy trade-offs under mitigation~\cite{chang2021,kulynych2022,suleymanov2026,tayebi2024}, and memorization in medical generative and foundation models~\cite{dar2025,parampottupadam2026}. However, none of these studies relates the strength of demographic encoding to patient-level membership risk. We are also not aware of a study that measures all three quantities on the same models and tests their alignment directly.

We test whether demographic encoding, diagnostic fairness gaps and MIA vulnerability are aligned or distinct in chest X-ray foundation models. We find that they are distinct. Encoding strength does not predict which patients leak, erasing linearly decodable demographics changes age-related fairness gaps but not leakage, and leakage is governed by head capacity, training length and patient atypicality.

\begin{figure}[t]
\centering
\adjustbox{max width=\linewidth}{%
\begin{tikzpicture}[
  box/.style={draw=black!70, rounded corners=2pt, fill=#1, minimum height=0.9cm, align=center, font=\footnotesize, inner sep=3pt},
  box/.default=white, >=Stealth, every node/.style={font=\footnotesize}]
\node[box=blue!8] (enc) {Frozen encoder\\BiomedCLIP / DINOv2 / RAD-DINO};
\node[box=blue!8, right=0.9cm of enc] (emb) {Patient-disjoint\\embeddings\\(audit/val/test)};
\node[box=orange!12, right=0.9cm of emb] (heads) {200 heads (linear / MLP)\\each patient in exactly\\100 models (balanced)};
\node[box=green!10, right=0.9cm of heads] (lira) {Patient-level\\LiRA audit\\(max over records)};
\draw[->] (enc)--(emb); \draw[->] (emb)--(heads); \draw[->] (heads)--(lira);
\node[box=gray!10, below=0.9cm of emb]   (probe) {Encoding\\OOF probes:\\sex, race, age};
\node[box=gray!10, below=0.9cm of heads] (fair)  {Fairness\\FNR / underdiagnosis\\gaps (reference\\head, test set)};
\node[box=gray!10, below=0.9cm of lira]  (mia)   {Leakage\\MIA AUC, TPR at\\fixed FPR,\\patient tail};
\draw[->] (emb)--(probe); \draw[->] (heads)--(fair); \draw[->] (lira)--(mia);
\node[draw=black!70, dashed, rounded corners=2pt, fit=(probe)(fair)(mia), inner sep=4pt, label={[font=\footnotesize]below:{Analyses: partial correlation and equivalence tests $\cdot$ LEACE erasure $\cdot$ capacity sweep $\cdot$ patient-level drivers of leakage}}] {};
\end{tikzpicture}}
\caption{Study design. Frozen embeddings feed 200 heads, each trained on a random half of the audit patients. The same embeddings and heads yield three measurements (encoding, fairness and leakage), which are then related to each other. OOF: out-of-fold; LiRA: likelihood-ratio attack; FNR: false-negative rate; TPR: true-positive rate; FPR: false-positive rate.}
\label{fig:pipeline}
\end{figure}
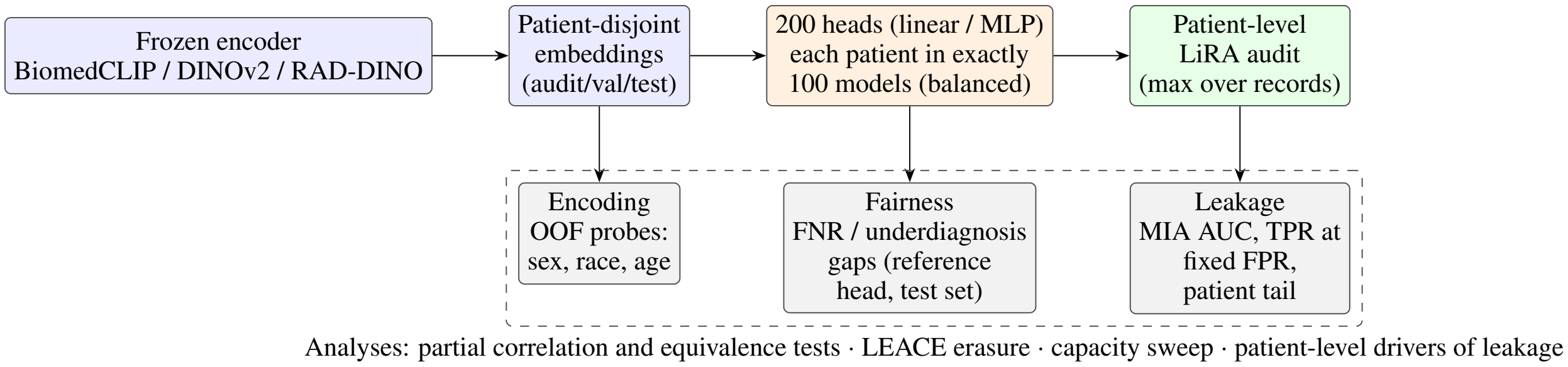

We make four contributions. First, we design a joint audit that measures demographic encoding, diagnostic fairness gaps and patient-level MIA vulnerability for the same 200-model ensembles, on two public chest X-ray datasets and three encoders, with three seeds for the main configurations (Section~\ref{sec:methods} and Figure~\ref{fig:pipeline}). Second, we use equivalence tests, rather than non-significant $p$-values, to show that encoding strength is statistically unrelated to leakage. All 72 CheXpert associations are equivalent to zero, and on NIH ChestX-ray14 the only failures are the six fully trained MLP runs for age, all with a negative sign (Section~\ref{sec:equiv}). Third, we provide intervention evidence with linear concept erasure. Erasure removes most linearly decodable demographic information without changing leakage, but it reduces age-related underdiagnosis gaps, which dissociates privacy from fairness (Section~\ref{sec:leace}). Fourth, we identify what drives leakage. Head capacity, training length, embedding family and patient atypicality explain leakage, whereas an apparent image-count effect is largely produced by taking the maximum over a patient's records (Sections~\ref{sec:capacity} and~\ref{sec:who}).

%=====================================================================
\section{Related work}\label{sec:related}

\subsection{Demographic encoding and fairness in chest radiography}
Underdiagnosis by chest X-ray classifiers is larger for female, young, Black and low-income patients~\cite{seyyedkalantari2021}. Deep models can predict self-reported race from images, even when the signal is hard to localize~\cite{gichoya2022}. Disease-detection models also encode protected characteristics in ways that relate to performance disparities~\cite{glocker2023}. Fairness gaps need not transfer across sites~\cite{yang2024}, and removing demographic information from representations does not guarantee fair behavior~\cite{petersen2023}. The studies closest to our encoding analysis are those of Kim et al.~\cite{kim2026}, who contrast what a foundation model encodes with what its linear head uses, and Santos et al.~\cite{santos2025}, who show that demographic information persists in X-ray foundation embeddings. Neither study measures privacy.

\subsection{Membership inference and disparate vulnerability}
Membership inference asks whether a record was in a model's training set~\cite{shokri2017}. Likelihood-ratio attacks calibrate this inference at low false-positive rates~\cite{carlini2022}. Vulnerability differs across groups~\cite{kulynych2022}, and fairness interventions can change it~\cite{chang2021}. Knolle et al.~\cite{knolle2026} audited medical imaging models at scale with a closed-form record-level likelihood-ratio test. They found that risk concentrates in a tail of patients and rises with architecture capacity. They also observed that the over-representation of some groups in this tail is partly explained by group size. In medical generative and fine-tuned foundation models, memorization has been shown to depend on the rarity and distinctiveness of samples~\cite{dar2025,parampottupadam2026}. Fairness-privacy trade-offs under mitigation have been audited at the subpopulation level~\cite{suleymanov2026}, and private training can preserve or erode fairness~\cite{tayebi2024}. These studies relate privacy to group membership, mitigation or sample rarity. To our knowledge, none of them asks whether demographic encoding predicts patient-level leakage, which is the question we address.

\subsection{Concept erasure}
Iterative nullspace projection removes linearly decodable attribute information from representations~\cite{ravfogel2020}. Least-squares concept erasure (LEACE) performs this removal in closed form and guarantees that no linear classifier can recover the attribute on the data used to fit it~\cite{belrose2023}. We use LEACE as an intervention to test whether removing encoded demographic information changes leakage.

%=====================================================================
\section{Methods}\label{sec:methods}

\subsection{Data and encoders}\label{sec:data}
We used two public chest radiograph datasets (Table~\ref{tab:data}). CheXpert-small~\cite{irvin2019} provides sex, age group and self-reported race (race labels from the demographic file released by Stanford AIMI \cite{gichoya2022}), and NIH ChestX-ray14~\cite{wang2017} provides sex and age group only. From each dataset, we sampled 15{,}000 patients once (seed 0). For each run seed, we split them at the patient level into audit (10{,}500), validation (1{,}500) and test (3{,}000) sets. The age of a patient is the mean over the records of that patient, grouped into four bins: under 40, 40 to 60, 60 to 80, and 80 years or older, with lower bounds inclusive. CheXpert has 14 labels; we masked uncertain labels and treated blank labels as negative. NIH ChestX-ray14 has 15 labels.

Embeddings came from three frozen encoders. BiomedCLIP~\cite{zhang2023biomedclip} provides 512-dimensional embeddings from its image tower. DINOv2~\cite{oquab2023} provides 768-dimensional class-token (CLS) embeddings and serves as a general-domain control. RAD-DINO~\cite{perezgarcia2025} provides 768-dimensional CLS embeddings and is specific to radiology. On NIH ChestX-ray14, we used BiomedCLIP and RAD-DINO only (Table~\ref{tab:data}). The pretraining data of RAD-DINO include CheXpert. We therefore treat its CheXpert results as an upper bound on the behavior of a domain-matched encoder, not as a clean test of generalization. We standardized all embeddings using audit-set statistics.

\begin{table}[t]
\centering\small
\caption{Datasets and encoders. Image counts are those used after sampling 15{,}000 patients.}\label{tab:data}
\resizebox{\linewidth}{!}{%
\begin{tabular}{@{}lcccccl@{}}
\toprule
Dataset & Patients & Images & Audit records & Labels & Attributes & Encoders\\
\midrule
CheXpert-small & 15{,}000 & 44{,}622 & 31{,}601 & 14 & sex, race, age & BiomedCLIP, DINOv2, RAD-DINO\\
NIH ChestX-ray14 & 15{,}000 & 54{,}521 & 38{,}316 & 15 & sex, age & BiomedCLIP, RAD-DINO\\
\bottomrule
\end{tabular}}
\end{table}

\subsection{Heads and membership audit}\label{sec:audit}
Following Knolle et al.~\cite{knolle2026}, we trained $N=200$ models, each on a random half of the audit patients, so that every patient was a member of exactly 100 models. Each model was either a linear head or a one-hidden-layer multilayer perceptron (MLP) on the frozen embedding. The loss, the training settings and the metrics are described in Section~\ref{sec:metrics}.

Let $\mathbf{x}_i$ be the embedding of record $i$, $f_{mj}(\mathbf{x}_i)$ the logit of model $m$ for label $j$, $y_{ij}\in\{0,1\}$ the label and $v_{ij}\in\{0,1\}$ an indicator that the label is valid. The record-level score was the mean signed logit margin over valid labels,
\begin{equation}
s_{mi}=\frac{\sum_{j}v_{ij}\,(2y_{ij}-1)\,f_{mj}(\mathbf{x}_i)}{\sum_{j}v_{ij}}.
\label{eq:score}
\end{equation}
\noindent For each record $i$, we computed the closed-form likelihood-ratio attack (LiRA) statistic
\begin{equation}
\theta_i=\Phi\left(\frac{\mu_{\mathrm{in}}-\mu_{\mathrm{out}}}{\sqrt{\sigma^{2}_{\mathrm{in}}+\sigma^{2}_{\mathrm{out}}}}\right),
\label{eq:lira}
\end{equation}
\noindent where $\mu_{\mathrm{in}}$ and $\sigma^{2}_{\mathrm{in}}$ are the mean and variance of the record score $s_{mi}$ over the models that contain the record, $\mu_{\mathrm{out}}$ and $\sigma^{2}_{\mathrm{out}}$ are the same quantities over the models that exclude it (the record index is omitted for readability), and $\Phi$ is the standard normal cumulative distribution function. The statistic $\theta_i$ equals the area under the curve (AUC) that separates the in and out score distributions when both are Gaussian.

As a calibration check, we re-drew membership at random. This null yielded a patient-level 99th percentile of approximately 0.606 in every run, and the closed-form and empirical AUCs agreed. We also report two alternative patient scores that remove the effect of taking the maximum over records, namely the mean over records and the score of one random record.

\subsection{Loss, metrics and hyperparameters}\label{sec:metrics}
Each model minimized the binary cross-entropy over the valid labels of its own member records, averaged over labels and records. We drew minibatches of $B=2048$ records with replacement from all $R$ audit records and trained all $N$ models on the same minibatch, with the loss of each model restricted to its member records (about half of each minibatch). One epoch corresponds to $\lceil R/B\rceil$ steps. Table~\ref{tab:hyper} lists the hyperparameters of the heads and of the linear probes used for encoding (Section~\ref{sec:encoding}). Weights were initialized from a Gaussian distribution scaled by the inverse square root of the fan-in, and biases were set to zero. The reference head of Section~\ref{sec:fairness} used the same settings but was trained on all audit patients.

\begin{table}[t]
\centering\small
\caption{Hyperparameters of the heads and of the encoding probes of Section~\ref{sec:encoding}. One epoch is $\lceil R/B\rceil$ minibatch steps. The probes of Table~\ref{tab:leace} are described in Section~\ref{sec:sweeps}.}\label{tab:hyper}
\begin{tabular}{@{}llp{0.48\linewidth}@{}}
\toprule
 & Setting & Value\\
\midrule
Heads & Architecture & Linear, or MLP with GELU (512 hidden units; 8 and 64 in the sweep)\\
 & Number of models $N$ & 200 (each patient in 100 models)\\
 & Optimizer, weight decay & AdamW~\cite{loshchilov2019}, $10^{-2}$\\
 & Learning rate & $3\times10^{-3}$ (linear), $10^{-3}$ (MLP)\\
 & Schedule & Cosine annealing to zero\\
 & Batch size $B$, epochs & 2048, 30 (5 in the sweep)\\
 & Seeds & 3 (main configurations), 1 (sweep)\\
Probes & Optimizer, weight decay & AdamW, $10^{-4}$\\
 & Learning rate, batch size & $3\times10^{-3}$, 2048\\
 & Epochs, cross-validation & 30, 5 folds\\
\bottomrule
\end{tabular}
\end{table}

We measured utility as the macro-averaged area under the receiver operating characteristic curve (AUROC) over the labels with both classes present, computed for the reference head on the test patients.

Leakage was measured at the record, aggregate and patient levels. Let $z_{mi}\in\{0,1\}$ indicate whether record $i$ belongs to the training set of model $m$. For the aggregate attack, we computed for every pair of model $m$ and record $i$ the log-likelihood ratio
\begin{equation}
\Lambda_{mi}=\log\frac{\mathcal{N}\big(s_{mi};\mu_{\mathrm{in}}^{(-m)},\sigma_{\mathrm{in}}^{2\,(-m)}\big)}{\mathcal{N}\big(s_{mi};\mu_{\mathrm{out}}^{(-m)},\sigma_{\mathrm{out}}^{2\,(-m)}\big)},
\label{eq:llr}
\end{equation}
\noindent where $\mathcal{N}(\cdot;\mu,\sigma^{2})$ is the Gaussian density and the means and variances are estimated from the models other than $m$. The aggregate MIA AUC is the area under the curve that separates members ($z_{mi}=1$) from non-members ($z_{mi}=0$) by $\Lambda_{mi}$, computed on a random sample of at most four million pairs. We also report the true-positive rate (TPR) at a false-positive rate (FPR) of $\alpha=10^{-3}$,
\begin{equation}
\mathrm{TPR@}\alpha=\Pr\big(\Lambda_{mi}>\lambda_{\alpha}\mid z_{mi}=1\big),\quad\text{with}\quad\Pr\big(\Lambda_{mi}>\lambda_{\alpha}\mid z_{mi}=0\big)=\alpha,
\label{eq:tpr}
\end{equation}
\noindent where $\lambda_{\alpha}$ is the decision threshold. The record-level statistic $\theta_i$ was aggregated to the patient level by
\begin{equation}
\theta_p=\max_{i\in\mathcal{R}_p}\theta_i,
\label{eq:patient}
\end{equation}
\noindent where $\mathcal{R}_p$ is the set of records of patient $p$. We summarize $\theta_p$ by its 99th percentile and by the share of patients with $\theta_p\ge0.9$.

For an attribute with $C$ classes, the patient-level encoding score is the prior-corrected posterior probability of the true class $c_p$ of patient $p$,
\begin{equation}
e_p=\frac{\exp\big(o_{pc_p}-\log\pi_{c_p}\big)}{\sum_{c=1}^{C}\exp\big(o_{pc}-\log\pi_{c}\big)},
\label{eq:enc}
\end{equation}
\noindent where $o_{pc}$ is the out-of-fold logit of class $c$ from the linear probe and $\pi_c$ is the class prior in the training folds. The chance level of $e_p$ is $1/C$. The overall encoding strength is the one-vs-rest macro-averaged AUROC of the posteriors.

For a group $g$ of test patients, and with the per-label thresholds of Section~\ref{sec:fairness}, the macro false-negative rate (FNR) and the underdiagnosis rate are
\begin{equation}
\mathrm{FNR}_g=\frac{1}{|\mathcal{J}_g|}\sum_{j\in\mathcal{J}_g}\frac{\mathrm{FN}_{gj}}{\mathrm{FN}_{gj}+\mathrm{TP}_{gj}},\qquad
\mathrm{UD}_g=\frac{\mathrm{FP}_{g0}}{\mathrm{FP}_{g0}+\mathrm{TN}_{g0}},
\label{eq:rates}
\end{equation}
\noindent where $\mathrm{FN}$, $\mathrm{TP}$, $\mathrm{FP}$ and $\mathrm{TN}$ are record counts at these thresholds, $\mathcal{J}_g$ is the set of disease labels with at least 20 positive records in $g$, and index 0 denotes the label ``No Finding''. The underdiagnosis rate is therefore the record-level FPR of ``No Finding'', computed over records with a valid negative ``No Finding'' label. The gaps relative to the reference group $r$ are
\begin{equation}
\Delta^{\mathrm{FNR}}_g=\mathrm{FNR}_g-\mathrm{FNR}_r,\qquad
\Delta^{\mathrm{UD}}_g=\mathrm{UD}_g-\mathrm{UD}_r.
\label{eq:gaps}
\end{equation}

\subsection{Demographic encoding}\label{sec:encoding}
For each attribute, we fitted linear probes on patient-mean embeddings and obtained out-of-fold (OOF) predictions with 5-fold cross-validation. The encoding score and the encoding strength are defined in Eq.~\eqref{eq:enc} and Section~\ref{sec:metrics}. The probes used all 15{,}000 sampled patients, and the scores of the audit patients entered the association analysis. For CheXpert race, we excluded the categories ``Unknown'' and ``Other''.

\subsection{Diagnostic fairness}\label{sec:fairness}
We trained a reference head on all audit patients and fixed per-label thresholds on the validation set using the Youden index~\cite{youden1950}. On the held-out test patients, we computed the FNR gap and the underdiagnosis gap (Eq.~\eqref{eq:gaps}) relative to the largest group of each attribute, which served as the reference group. The underdiagnosis rate is the rate at which patients with disease are called healthy. Within each run, we estimated confidence intervals (CIs) of the gaps with a patient-clustered bootstrap of 500 resamples. For the differences between original and erased embeddings across seeds (Section~\ref{sec:leace}), we used a $t$-interval over the three seeds. The Results focus on the underdiagnosis gap, which is the primary measure of Seyyed-Kalantari et al.~\cite{seyyedkalantari2021}. The FNR gap was computed with the same procedure but is not analyzed further.

\subsection{Association and equivalence analysis}\label{sec:assoc}
We related patient encoding scores to patient MIA vulnerability in two ways. The first was the rank-transformed partial correlation $\rho$. The second was the increment $\Delta R^2$ that encoding adds to a regression that also contains log record count, $k$-nearest-neighbor sparsity ($k=10$, cosine distance), number of findings and log group size. A non-significant $p$-value does not show that an association is absent, so we used equivalence tests~\cite{schuirmann1987,lakens2017}. We call an association equivalent to zero if two conditions hold. The 95\% CI of $\rho$ must lie within $\pm0.1$, which is a conservative two-one-sided-test criterion, and the upper bound of the CI of $\Delta R^2$ must be below 0.01. All variables were rank-transformed. Sparsity was computed on patient-mean embeddings within the audit set, the number of findings is the mean number of positive findings per record, and log group size is the logarithm of the number of audit patients in the group of the patient. The CIs of $\rho$ and $\Delta R^2$ are percentile intervals from a patient-level bootstrap of 500 resamples.

For the analysis of Section~\ref{sec:who}, we fitted a broader rank regression with eight predictor blocks on the seed-0 runs. The blocks were log record count, sparsity, number of findings, the mean fraction of uncertain labels per record, the mean number of rare findings per record (labels with a prevalence below 5\%), the fraction of records with support devices, patient age, and demographic encoding (the mean of the attribute encoding scores). For each block, we report the drop-one $\Delta R^2$, that is, the decrease in $R^2$ when the block is removed from the full model. Because the blocks are correlated, these values do not sum to the total $R^2$.

\subsection{Interventions and sweeps}\label{sec:sweeps}
We applied LEACE~\cite{belrose2023} to erase the one-hot sex, race and age concepts jointly on CheXpert and then repeated the audit. We fitted LEACE at the record level on all records of the 15{,}000 sampled patients, including validation and test patients, with one-hot columns for all observed race categories. Because LEACE serves here as an analytical intervention rather than a deployable mitigation, this transductive fit applies the same erasure map to every audited and evaluated record. To measure residual demographic information (Table~\ref{tab:leace}), we fitted separate 5-fold probes on re-standardized patient-mean embeddings of all sampled patients. These were a linear probe and an MLP probe with one hidden layer of 256 ReLU units, trained with class-weighted cross-entropy, AdamW (learning rate $10^{-3}$, weight decay $10^{-2}$), 20 epochs and a batch size of 512. They report the macro-AUROC of uncorrected posteriors, so their values differ from the encoding strengths of Section~\ref{sec:equiv}, which use the probes of Section~\ref{sec:encoding}.

To study capacity and training length, we varied the MLP width (8, 64 and 512 hidden units) and the number of epochs (5 and 30) for BiomedCLIP and RAD-DINO on seed 0. On NIH ChestX-ray14, the sweep included only the 5-epoch MLP. Table~\ref{tab:equiv} pools these sweep runs under ``MLP reduced''. The main configurations (linear and 30-epoch MLP heads) used three seeds, and each seed re-drew the patient split, the membership assignment and the head initialization. We report the mean $\pm$ standard deviation (s.d.).

%=====================================================================
\section{Results}\label{sec:results}

\subsection{Leakage depends on encoder and head, not on utility}\label{sec:leakage}
\begin{figure}[t]
\centering
\resizebox{\linewidth}{!}{\input{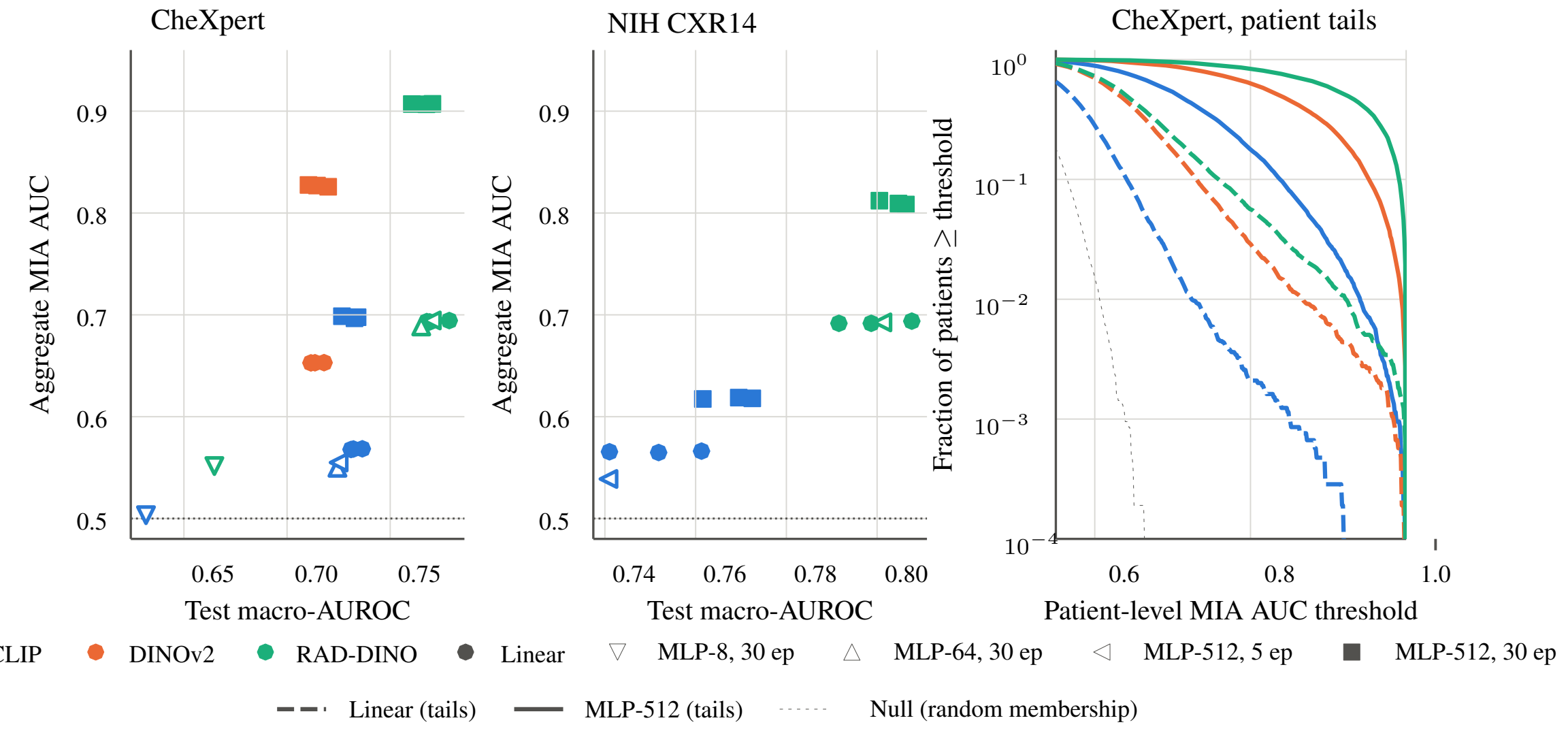}}
\caption{Leakage is decoupled from utility. (a, b) Aggregate MIA AUC versus test AUROC on CheXpert and NIH ChestX-ray14. (c) Survival curve of patient-level membership AUC on CheXpert; the gray curve shows the null with random membership.}\label{fig:leak}
\end{figure}

Leakage varied widely while test performance changed little (Table~\ref{tab:main} and Figure~\ref{fig:leak}). On CheXpert, a 30-epoch MLP on RAD-DINO reached an aggregate MIA AUC of 0.907. It placed 59.6\% of patients above an AUC of 0.9 and achieved a TPR of 12.4\% at an FPR of $10^{-3}$. A linear head on BiomedCLIP gave corresponding values of 0.568, 0.0\% and 0.5\%. Its test AUROC was modestly lower, at 0.733 compared with 0.766 for the RAD-DINO MLP and 0.772 for the RAD-DINO linear head.

For all three encoders, the MLP head leaked more than the linear head without a gain in test AUROC. The MIA AUC rose from 0.568 to 0.698 for BiomedCLIP, from 0.653 to 0.827 for DINOv2 and from 0.694 to 0.907 for RAD-DINO. The pattern replicated on NIH ChestX-ray14, where the MIA AUC rose from 0.692 to 0.810 for RAD-DINO and from 0.566 to 0.618 for BiomedCLIP. This result is consistent with the capacity effect reported for end-to-end models~\cite{knolle2026} and shows the same effect for small heads on frozen features. Variation across seeds was negligible, with an s.d.\ of at most 0.002 in MIA AUC.

\begin{table}[t]
\centering\scriptsize
\caption{Utility and leakage (mean $\pm$ s.d. over three seeds). The column Patients AUC$\ge$0.9 gives the share of patients whose max-over-records membership AUC is at least 0.9.}\label{tab:main}
\resizebox{\linewidth}{!}{\begin{tabular}{@{}llccccc@{}}
\toprule
Dataset & Encoder / head & Test AUROC & MIA AUC & TPR@$10^{-3}$FPR (\%) & Patient AUC p99 & Patients AUC$\ge$0.9 (\%) \\
\midrule
CheXpert & BiomedCLIP / Linear & 0.733\,$\pm$\,0.003 & 0.568\,$\pm$\,0.000 & 0.47\,$\pm$\,0.04 & 0.725\,$\pm$\,0.001 & 0.0\,$\pm$\,0.0 \\
 & BiomedCLIP / MLP & 0.731\,$\pm$\,0.004 & 0.698\,$\pm$\,0.001 & 1.62\,$\pm$\,0.06 & 0.939\,$\pm$\,0.003 & 2.9\,$\pm$\,0.0 \\
 & DINOv2 / Linear & 0.714\,$\pm$\,0.003 & 0.653\,$\pm$\,0.000 & 1.37\,$\pm$\,0.05 & 0.869\,$\pm$\,0.005 & 0.6\,$\pm$\,0.0 \\
 & DINOv2 / MLP & 0.715\,$\pm$\,0.005 & 0.827\,$\pm$\,0.001 & 5.34\,$\pm$\,0.09 & 0.993\,$\pm$\,0.001 & 28.3\,$\pm$\,0.1 \\
 & RAD-DINO / Linear & 0.772\,$\pm$\,0.006 & 0.694\,$\pm$\,0.001 & 2.88\,$\pm$\,0.12 & 0.918\,$\pm$\,0.004 & 1.4\,$\pm$\,0.1 \\
 & RAD-DINO / MLP & 0.766\,$\pm$\,0.005 & 0.907\,$\pm$\,0.000 & 12.43\,$\pm$\,0.22 & 1.000\,$\pm$\,0.000 & 59.6\,$\pm$\,0.4 \\
\midrule
NIH CXR14 & BiomedCLIP / Linear & 0.751\,$\pm$\,0.010 & 0.566\,$\pm$\,0.001 & 0.94\,$\pm$\,0.10 & 0.725\,$\pm$\,0.005 & 0.1\,$\pm$\,0.0 \\
 & BiomedCLIP / MLP & 0.768\,$\pm$\,0.006 & 0.618\,$\pm$\,0.001 & 1.30\,$\pm$\,0.02 & 0.908\,$\pm$\,0.001 & 1.1\,$\pm$\,0.0 \\
 & RAD-DINO / Linear & 0.799\,$\pm$\,0.008 & 0.692\,$\pm$\,0.001 & 4.80\,$\pm$\,0.25 & 0.932\,$\pm$\,0.004 & 1.5\,$\pm$\,0.0 \\
 & RAD-DINO / MLP & 0.804\,$\pm$\,0.003 & 0.810\,$\pm$\,0.002 & 7.16\,$\pm$\,0.09 & 0.998\,$\pm$\,0.000 & 16.8\,$\pm$\,0.3 \\
\bottomrule
\end{tabular}}
\end{table}

\subsection{Training length and capacity control leakage}\label{sec:capacity}
Leakage is not a property of the encoder alone (Table~\ref{tab:regime}; all values in this subsection are from seed 0). With RAD-DINO on CheXpert, a 512-unit MLP trained for 5 epochs matched the MIA AUC of the linear head (0.694) at a test AUROC of 0.770. Training the same MLP for 30 epochs raised the MIA AUC to 0.907 without a utility gain. A narrow MLP with 8 units leaked little (0.552) but lost utility (test AUROC of 0.665). An MLP with 64 units already leaked about as much as the linear head (0.689). BiomedCLIP showed the same shape, with the 512-unit MLP reaching an MIA AUC of 0.555 after 5 epochs and 0.698 after 30 epochs, at test AUROCs of 0.725 and 0.734, respectively. On NIH ChestX-ray14, the MIA AUC of RAD-DINO rose from 0.693 to 0.808. Shorter training was not always free on this dataset: for BiomedCLIP, the 5-epoch MLP lost about 0.02 in test AUROC relative to the 30-epoch MLP (0.741 versus 0.762). The 5-epoch setting is a fixed proxy for early stopping and does not use validation-based early stopping.

\begin{table}[t]
\centering\scriptsize
\vspace{0.8pt}
\caption{Capacity and training-length sweep over MLP width and number of epochs (seed 0).}\label{tab:regime}
\begin{tabular}{@{}llcccc@{}}
\toprule
Encoder & Head & Test AUROC & MIA AUC & TPR@$10^{-3}$FPR (\%) & Patients AUC$\ge$0.9 (\%) \\
\midrule
CheXpert / BiomedCLIP & Linear & 0.736 & 0.568 & 0.51 & 0.0 \\
 & MLP-8, 30 ep & 0.632 & 0.503 & 0.11 & 0.0 \\
 & MLP-64, 30 ep & 0.724 & 0.550 & 0.24 & 0.0 \\
 & MLP-512, 5 ep & 0.725 & 0.555 & 0.35 & 0.0 \\
 & MLP-512, 30 ep & 0.734 & 0.698 & 1.67 & 2.9 \\
\midrule
CheXpert / RAD-DINO & Linear & 0.779 & 0.694 & 3.02 & 1.4 \\
 & MLP-8, 30 ep & 0.665 & 0.552 & 0.41 & 0.0 \\
 & MLP-64, 30 ep & 0.765 & 0.689 & 1.54 & 1.0 \\
 & MLP-512, 5 ep & 0.770 & 0.694 & 2.46 & 2.9 \\
 & MLP-512, 30 ep & 0.770 & 0.907 & 12.63 & 59.3 \\
\midrule
NIH CXR14 / BiomedCLIP & Linear & 0.752 & 0.565 & 0.82 & 0.1 \\
 & MLP-512, 5 ep & 0.741 & 0.539 & 0.37 & 0.0 \\
 & MLP-512, 30 ep & 0.762 & 0.617 & 1.28 & 1.2 \\
\midrule
NIH CXR14 / RAD-DINO & Linear & 0.799 & 0.692 & 4.58 & 1.4 \\
 & MLP-512, 5 ep & 0.801 & 0.693 & 4.02 & 2.8 \\
 & MLP-512, 30 ep & 0.806 & 0.808 & 7.05 & 16.9 \\
\bottomrule
\end{tabular}
\end{table}

\subsection{Demographic encoding is statistically unrelated to patient vulnerability}\label{sec:equiv}
All three encoders encoded demographics linearly. On CheXpert, the encoding macro-AUROC for age, race and sex was 0.78, 0.68 and 0.89 for BiomedCLIP, 0.85, 0.76 and 0.98 for DINOv2, and 0.90, 0.78 and 1.00 for RAD-DINO. On NIH ChestX-ray14, the values for age and sex were 0.72 and 0.94 for BiomedCLIP and 0.87 and 1.00 for RAD-DINO.

\begin{figure}[t]
\centering
\resizebox{\linewidth}{!}{\input{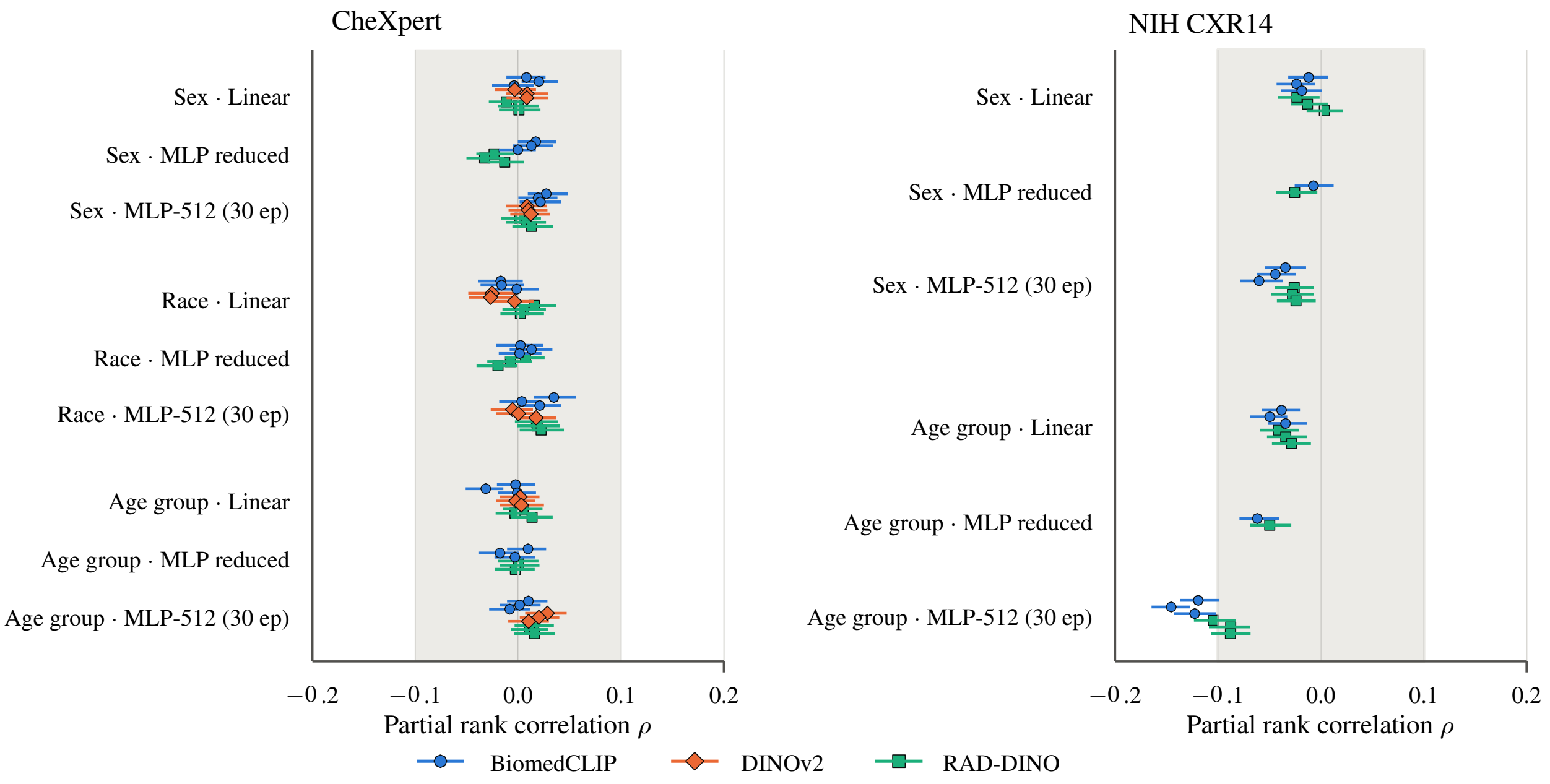}}
\caption{Partial rank correlation between demographic encoding and membership vulnerability. Points are individual runs with 95\% CIs, and the shaded band is the equivalence region of $\pm0.1$.}\label{fig:equiv}
\end{figure}

However, within a model, the strength with which the demographics of a patient were encoded did not predict the vulnerability of that patient (Figure~\ref{fig:equiv} and Table~\ref{tab:equiv}). On CheXpert, all 72 associations (attributes $\times$ head regimes $\times$ encoders $\times$ seeds) were equivalent to zero. Their absolute partial correlations were at most 0.035, the widest 95\% CI was $\pm0.056$, and the largest upper bound on $\Delta R^2$ was 0.0018. On NIH ChestX-ray14, 22 of 28 associations passed the $\rho$ criterion and 27 of 28 passed the $\Delta R^2$ criterion. All six exceptions to the $\rho$ criterion were fully trained MLP runs for age, with $\rho$ between $-0.145$ and $-0.088$, and the single exception to the $\Delta R^2$ criterion (upper bound 0.0114) occurred in the same group of runs. Their sign was negative, so stronger age encoding went with lower leakage. For these runs, we therefore claim only that encoding strength did not increase leakage, not that the association was equivalent to zero.

At the model level, encoding strength and leakage gave the same ranking of encoders, with RAD-DINO first, DINOv2 second and BiomedCLIP third. This observation is suggestive but not conclusive, because it rests on only three encoders and the patient-level association is null. We interpret it as a possible shared driver, such as embedding richness, rather than as evidence of causation.

\begin{table}[t]
\centering\scriptsize
\caption{Equivalence tests by dataset, head regime and attribute. For the linear and 30-epoch MLP regimes, the column Runs counts encoder and seed combinations. ``MLP reduced'' pools the seed-0 sweep runs of BiomedCLIP and RAD-DINO (8- and 64-unit MLPs and the 5-epoch 512-unit MLP on CheXpert; the 5-epoch MLP on NIH ChestX-ray14), so Runs counts encoder and configuration combinations.}\label{tab:equiv}
\resizebox{\linewidth}{!}{\begin{tabular}{@{}llcccccc@{}}
\toprule
Dataset & Head regime / attribute & Runs & Partial $\rho$ range & Widest 95\% CI & $\rho$-equiv. & max $\Delta R^2$ (CI upper) & $\Delta R^2$-equiv. \\
\midrule
CheXpert & Linear / Sex & 9 & [-0.012, +0.020] & [-0.029, +0.039] & 9/9 & 0.0009 & 9/9 \\
 & Linear / Race & 9 & [-0.027, +0.016] & [-0.049, +0.037] & 9/9 & 0.0013 & 9/9 \\
 & Linear / Age group & 9 & [-0.032, +0.013] & [-0.051, +0.034] & 9/9 & 0.0016 & 9/9 \\
 & MLP-512, 30 ep / Sex & 9 & [+0.001, +0.027] & [-0.016, +0.048] & 9/9 & 0.0014 & 9/9 \\
 & MLP-512, 30 ep / Race & 9 & [-0.005, +0.035] & [-0.027, +0.056] & 9/9 & 0.0018 & 9/9 \\
 & MLP-512, 30 ep / Age group & 9 & [-0.008, +0.028] & [-0.028, +0.047] & 9/9 & 0.0011 & 9/9 \\
 & MLP reduced / Sex & 6 & [-0.033, +0.017] & [-0.050, +0.037] & 6/6 & 0.0015 & 6/6 \\
 & MLP reduced / Race & 6 & [-0.020, +0.013] & [-0.041, +0.033] & 6/6 & 0.0012 & 6/6 \\
 & MLP reduced / Age group & 6 & [-0.018, +0.010] & [-0.038, +0.027] & 6/6 & 0.0009 & 6/6 \\
\midrule
NIH CXR14 & Linear / Sex & 6 & [-0.024, +0.003] & [-0.043, +0.022] & 6/6 & 0.0012 & 6/6 \\
 & Linear / Age group & 6 & [-0.050, -0.029] & [-0.069, -0.010] & 6/6 & 0.0030 & 6/6 \\
 & MLP-512, 30 ep / Sex & 6 & [-0.060, -0.024] & [-0.078, -0.005] & 6/6 & 0.0027 & 6/6 \\
 & MLP-512, 30 ep / Age group & 6 & [-0.145, -0.088] & [-0.165, -0.068] & 0/6 & 0.0114 & 5/6 \\
 & MLP reduced / Sex & 2 & [-0.025, -0.007] & [-0.044, +0.012] & 2/2 & 0.0010 & 2/2 \\
 & MLP reduced / Age group & 2 & [-0.062, -0.050] & [-0.079, -0.029] & 2/2 & 0.0039 & 2/2 \\
\bottomrule
\end{tabular}}
\end{table}

\subsection{Patient atypicality, not demographics, predicts vulnerability}\label{sec:who}
\begin{figure}[t]
\centering
\resizebox{\linewidth}{!}{\begin{tikzpicture}[x=1cm,y=1cm,font=\fontsize{7}{8.4}\selectfont]
\definecolor{hlo}{RGB}{247,251,255}\definecolor{hhi}{RGB}{8,48,107}
\tikzset{cell/.style={minimum width=1.05cm,minimum height=0.9cm,inner sep=0,anchor=center}}
\node[font=\bfseries\fontsize{8.5}{10}\selectfont] at (1.57,0.9) {Max over records};
\node[cell,fill=hhi!85!hlo,text=white] at (0.00,0.00) {0.250};
\node[cell,fill=hhi!73!hlo,text=white] at (1.05,0.00) {0.192};
\node[cell,fill=hhi!100!hlo,text=white] at (2.10,0.00) {0.327};
\node[cell,fill=hhi!74!hlo,text=white] at (3.15,0.00) {0.196};
\node[cell,fill=hhi!5!hlo,text=black] at (0.00,-0.90) {0.002};
\node[cell,fill=hhi!25!hlo,text=black] at (1.05,-0.90) {0.032};
\node[cell,fill=hhi!3!hlo,text=black] at (2.10,-0.90) {0.001};
\node[cell,fill=hhi!9!hlo,text=black] at (3.15,-0.90) {0.006};
\node[cell,fill=hhi!8!hlo,text=black] at (0.00,-1.80) {0.005};
\node[cell,fill=hhi!24!hlo,text=black] at (1.05,-1.80) {0.030};
\node[cell,fill=hhi!16!hlo,text=black] at (2.10,-1.80) {0.016};
\node[cell,fill=hhi!40!hlo,text=black] at (3.15,-1.80) {0.072};
\node[cell,fill=hhi!3!hlo,text=black] at (0.00,-2.70) {0.001};
\node[cell,fill=hhi!7!hlo,text=black] at (1.05,-2.70) {0.004};
\node[cell,fill=hhi!3!hlo,text=black] at (2.10,-2.70) {0.001};
\node[cell,fill=hhi!3!hlo,text=black] at (3.15,-2.70) {0.001};
\node[cell,fill=hhi!11!hlo,text=black] at (0.00,-3.60) {0.008};
\node[cell,fill=hhi!36!hlo,text=black] at (1.05,-3.60) {0.059};
\node[cell,fill=hhi!14!hlo,text=black] at (2.10,-3.60) {0.012};
\node[cell,fill=hhi!12!hlo,text=black] at (3.15,-3.60) {0.009};
\node[cell,fill=hhi!5!hlo,text=black] at (0.00,-4.50) {0.002};
\node[cell,fill=hhi!13!hlo,text=black] at (1.05,-4.50) {0.011};
\node[cell,fill=hhi!10!hlo,text=black] at (2.10,-4.50) {0.007};
\node[cell,fill=hhi!13!hlo,text=black] at (3.15,-4.50) {0.011};
\node[cell,fill=hhi!0!hlo,text=black] at (0.00,-5.40) {0.000};
\node[cell,fill=hhi!0!hlo,text=black] at (1.05,-5.40) {0.000};
\node[cell,fill=hhi!0!hlo,text=black] at (2.10,-5.40) {0.000};
\node[cell,fill=hhi!5!hlo,text=black] at (3.15,-5.40) {0.002};
\node[cell,fill=hhi!0!hlo,text=black] at (0.00,-6.30) {0.000};
\node[cell,fill=hhi!0!hlo,text=black] at (1.05,-6.30) {0.000};
\node[cell,fill=hhi!0!hlo,text=black] at (2.10,-6.30) {0.000};
\node[cell,fill=hhi!0!hlo,text=black] at (3.15,-6.30) {0.000};
\node[align=center,text=black!70,font=\fontsize{7}{8.4}\selectfont] at (0.00,-7.45) {BC\\linear};
\node[align=center,text=black!70,font=\fontsize{7}{8.4}\selectfont] at (1.05,-7.45) {BC\\MLP};
\node[align=center,text=black!70,font=\fontsize{7}{8.4}\selectfont] at (2.10,-7.45) {RD\\linear};
\node[align=center,text=black!70,font=\fontsize{7}{8.4}\selectfont] at (3.15,-7.45) {RD\\MLP};
\node[font=\fontsize{7}{8.4}\selectfont] at (1.57,-8.2) {Total $R^2$: 0.38, 0.51, 0.53, 0.54};
\node[font=\bfseries\fontsize{8.5}{10}\selectfont] at (6.28,0.9) {Mean over records};
\node[cell,fill=hhi!28!hlo,text=black] at (4.70,0.00) {0.039};
\node[cell,fill=hhi!12!hlo,text=black] at (5.75,0.00) {0.010};
\node[cell,fill=hhi!61!hlo,text=black] at (6.80,0.00) {0.143};
\node[cell,fill=hhi!5!hlo,text=black] at (7.85,0.00) {0.002};
\node[cell,fill=hhi!3!hlo,text=black] at (4.70,-0.90) {0.001};
\node[cell,fill=hhi!34!hlo,text=black] at (5.75,-0.90) {0.054};
\node[cell,fill=hhi!6!hlo,text=black] at (6.80,-0.90) {0.003};
\node[cell,fill=hhi!14!hlo,text=black] at (7.85,-0.90) {0.012};
\node[cell,fill=hhi!10!hlo,text=black] at (4.70,-1.80) {0.007};
\node[cell,fill=hhi!31!hlo,text=black] at (5.75,-1.80) {0.046};
\node[cell,fill=hhi!20!hlo,text=black] at (6.80,-1.80) {0.022};
\node[cell,fill=hhi!56!hlo,text=black] at (7.85,-1.80) {0.124};
\node[cell,fill=hhi!3!hlo,text=black] at (4.70,-2.70) {0.001};
\node[cell,fill=hhi!12!hlo,text=black] at (5.75,-2.70) {0.009};
\node[cell,fill=hhi!5!hlo,text=black] at (6.80,-2.70) {0.002};
\node[cell,fill=hhi!7!hlo,text=black] at (7.85,-2.70) {0.004};
\node[cell,fill=hhi!14!hlo,text=black] at (4.70,-3.60) {0.012};
\node[cell,fill=hhi!38!hlo,text=black] at (5.75,-3.60) {0.064};
\node[cell,fill=hhi!16!hlo,text=black] at (6.80,-3.60) {0.016};
\node[cell,fill=hhi!8!hlo,text=black] at (7.85,-3.60) {0.005};
\node[cell,fill=hhi!5!hlo,text=black] at (4.70,-4.50) {0.002};
\node[cell,fill=hhi!18!hlo,text=black] at (5.75,-4.50) {0.019};
\node[cell,fill=hhi!12!hlo,text=black] at (6.80,-4.50) {0.010};
\node[cell,fill=hhi!19!hlo,text=black] at (7.85,-4.50) {0.021};
\node[cell,fill=hhi!0!hlo,text=black] at (4.70,-5.40) {0.000};
\node[cell,fill=hhi!0!hlo,text=black] at (5.75,-5.40) {0.000};
\node[cell,fill=hhi!0!hlo,text=black] at (6.80,-5.40) {0.000};
\node[cell,fill=hhi!7!hlo,text=black] at (7.85,-5.40) {0.004};
\node[cell,fill=hhi!0!hlo,text=black] at (4.70,-6.30) {0.000};
\node[cell,fill=hhi!0!hlo,text=black] at (5.75,-6.30) {0.000};
\node[cell,fill=hhi!0!hlo,text=black] at (6.80,-6.30) {0.000};
\node[cell,fill=hhi!0!hlo,text=black] at (7.85,-6.30) {0.000};
\node[align=center,text=black!70,font=\fontsize{7}{8.4}\selectfont] at (4.70,-7.45) {BC\\linear};
\node[align=center,text=black!70,font=\fontsize{7}{8.4}\selectfont] at (5.75,-7.45) {BC\\MLP};
\node[align=center,text=black!70,font=\fontsize{7}{8.4}\selectfont] at (6.80,-7.45) {RD\\linear};
\node[align=center,text=black!70,font=\fontsize{7}{8.4}\selectfont] at (7.85,-7.45) {RD\\MLP};
\node[font=\fontsize{7}{8.4}\selectfont] at (6.28,-8.2) {Total $R^2$: 0.10, 0.28, 0.29, 0.27};
\node[font=\bfseries\fontsize{8.5}{10}\selectfont] at (10.97,0.9) {One random record};
\node[cell,fill=hhi!21!hlo,text=black] at (9.40,0.00) {0.025};
\node[cell,fill=hhi!6!hlo,text=black] at (10.45,0.00) {0.003};
\node[cell,fill=hhi!51!hlo,text=black] at (11.50,0.00) {0.108};
\node[cell,fill=hhi!8!hlo,text=black] at (12.55,0.00) {0.005};
\node[cell,fill=hhi!3!hlo,text=black] at (9.40,-0.90) {0.001};
\node[cell,fill=hhi!29!hlo,text=black] at (10.45,-0.90) {0.041};
\node[cell,fill=hhi!3!hlo,text=black] at (11.50,-0.90) {0.001};
\node[cell,fill=hhi!12!hlo,text=black] at (12.55,-0.90) {0.009};
\node[cell,fill=hhi!8!hlo,text=black] at (9.40,-1.80) {0.005};
\node[cell,fill=hhi!27!hlo,text=black] at (10.45,-1.80) {0.037};
\node[cell,fill=hhi!17!hlo,text=black] at (11.50,-1.80) {0.017};
\node[cell,fill=hhi!47!hlo,text=black] at (12.55,-1.80) {0.094};
\node[cell,fill=hhi!3!hlo,text=black] at (9.40,-2.70) {0.001};
\node[cell,fill=hhi!10!hlo,text=black] at (10.45,-2.70) {0.007};
\node[cell,fill=hhi!5!hlo,text=black] at (11.50,-2.70) {0.002};
\node[cell,fill=hhi!6!hlo,text=black] at (12.55,-2.70) {0.003};
\node[cell,fill=hhi!11!hlo,text=black] at (9.40,-3.60) {0.008};
\node[cell,fill=hhi!30!hlo,text=black] at (10.45,-3.60) {0.043};
\node[cell,fill=hhi!14!hlo,text=black] at (11.50,-3.60) {0.012};
\node[cell,fill=hhi!7!hlo,text=black] at (12.55,-3.60) {0.004};
\node[cell,fill=hhi!5!hlo,text=black] at (9.40,-4.50) {0.002};
\node[cell,fill=hhi!15!hlo,text=black] at (10.45,-4.50) {0.014};
\node[cell,fill=hhi!11!hlo,text=black] at (11.50,-4.50) {0.008};
\node[cell,fill=hhi!16!hlo,text=black] at (12.55,-4.50) {0.015};
\node[cell,fill=hhi!0!hlo,text=black] at (9.40,-5.40) {0.000};
\node[cell,fill=hhi!0!hlo,text=black] at (10.45,-5.40) {0.000};
\node[cell,fill=hhi!0!hlo,text=black] at (11.50,-5.40) {0.000};
\node[cell,fill=hhi!7!hlo,text=black] at (12.55,-5.40) {0.004};
\node[cell,fill=hhi!0!hlo,text=black] at (9.40,-6.30) {0.000};
\node[cell,fill=hhi!0!hlo,text=black] at (10.45,-6.30) {0.000};
\node[cell,fill=hhi!0!hlo,text=black] at (11.50,-6.30) {0.000};
\node[cell,fill=hhi!0!hlo,text=black] at (12.55,-6.30) {0.000};
\node[align=center,text=black!70,font=\fontsize{7}{8.4}\selectfont] at (9.40,-7.45) {BC\\linear};
\node[align=center,text=black!70,font=\fontsize{7}{8.4}\selectfont] at (10.45,-7.45) {BC\\MLP};
\node[align=center,text=black!70,font=\fontsize{7}{8.4}\selectfont] at (11.50,-7.45) {RD\\linear};
\node[align=center,text=black!70,font=\fontsize{7}{8.4}\selectfont] at (12.55,-7.45) {RD\\MLP};
\node[font=\fontsize{7}{8.4}\selectfont] at (10.97,-8.2) {Total $R^2$: 0.07, 0.20, 0.22, 0.22};
\node[anchor=east,text=black!70] at (-0.6,0.00) {Records per patient (log)};
\node[anchor=east,text=black!70] at (-0.6,-0.90) {kNN sparsity};
\node[anchor=east,text=black!70] at (-0.6,-1.80) {Number of findings};
\node[anchor=east,text=black!70] at (-0.6,-2.70) {Uncertain-label fraction};
\node[anchor=east,text=black!70] at (-0.6,-3.60) {Rare findings};
\node[anchor=east,text=black!70] at (-0.6,-4.50) {Support devices};
\node[anchor=east,text=black!70] at (-0.6,-5.40) {Age};
\node[anchor=east,text=black!70] at (-0.6,-6.30) {Demographic encoding};
\end{tikzpicture}}
\caption{Incremental $R^2$ of each feature block for patient vulnerability on CheXpert under three patient scores (maximum over records, mean over records and one random record). Cells show the drop-one $\Delta R^2$ of each block for the seed-0 runs, and Total $R^2$ is the $R^2$ of the full model. BC: BiomedCLIP; RD: RAD-DINO.}\label{fig:who}
\end{figure}
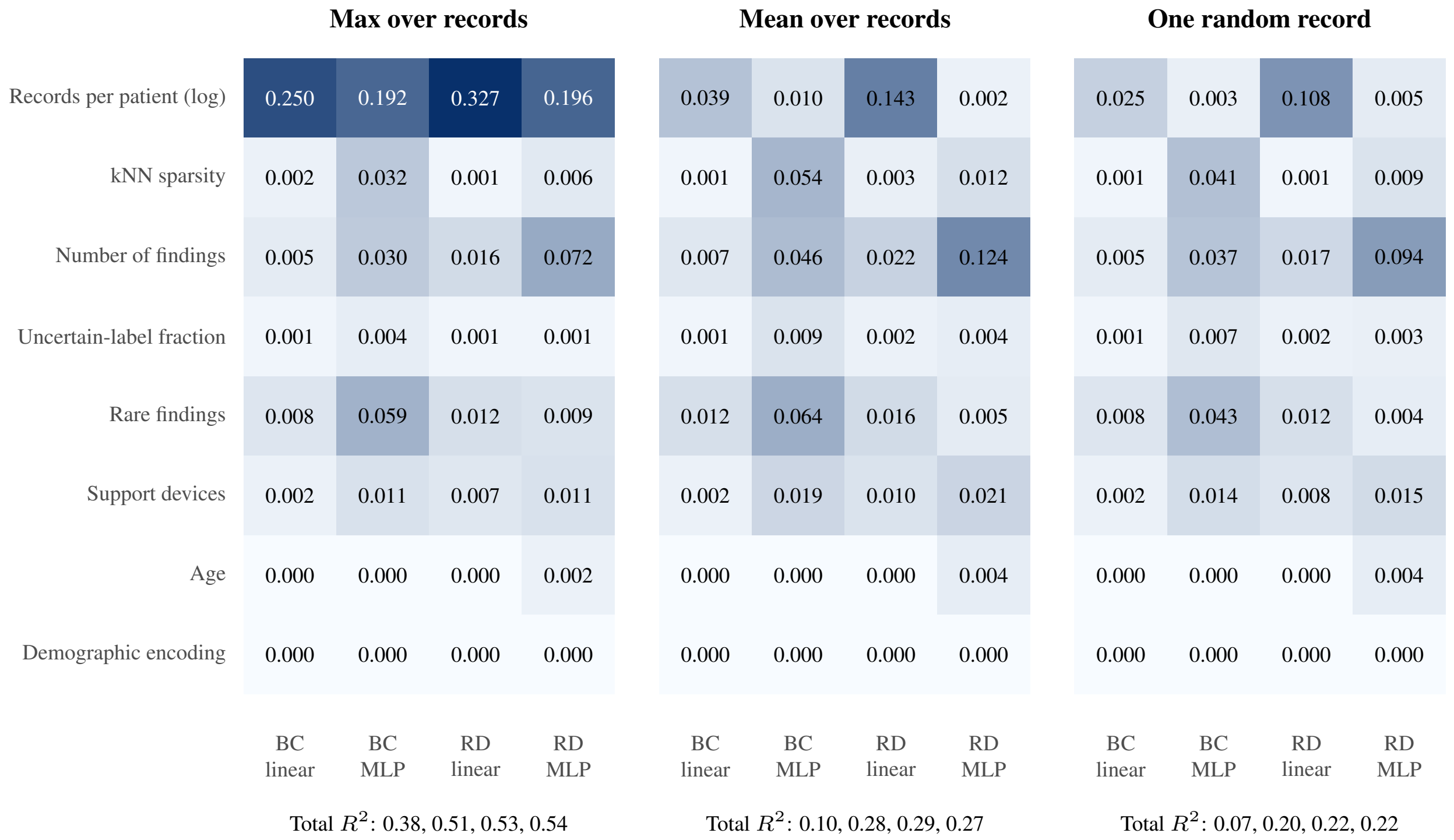

Under the max-over-records score, log record count alone explained a $\Delta R^2$ of 0.19 to 0.33 of patient vulnerability (Figure~\ref{fig:who}). This effect largely disappeared when patients were scored by the mean over records or by one random record ($\Delta R^2\le0.04$). The exception was the RAD-DINO linear head, with values of 0.143 and 0.108, respectively. The record-count effect therefore mainly reflects taking a maximum over more draws. The remaining explained variance came from features of patient atypicality. These features were the number of findings (up to 0.124 for the RAD-DINO MLP), rare findings and sparse embedding neighborhoods (up to 0.064 for the BiomedCLIP MLP), and support devices (at most 0.021). Demographic encoding contributed approximately zero under every score. The total $R^2$ under the alternative scores was 0.07 to 0.30, so these features leave most of the patient-level variance unexplained.

\subsection{Erasing demographics changes fairness, not leakage}\label{sec:leace}
\begin{figure}[t]
\centering
\resizebox{\linewidth}{!}{\input{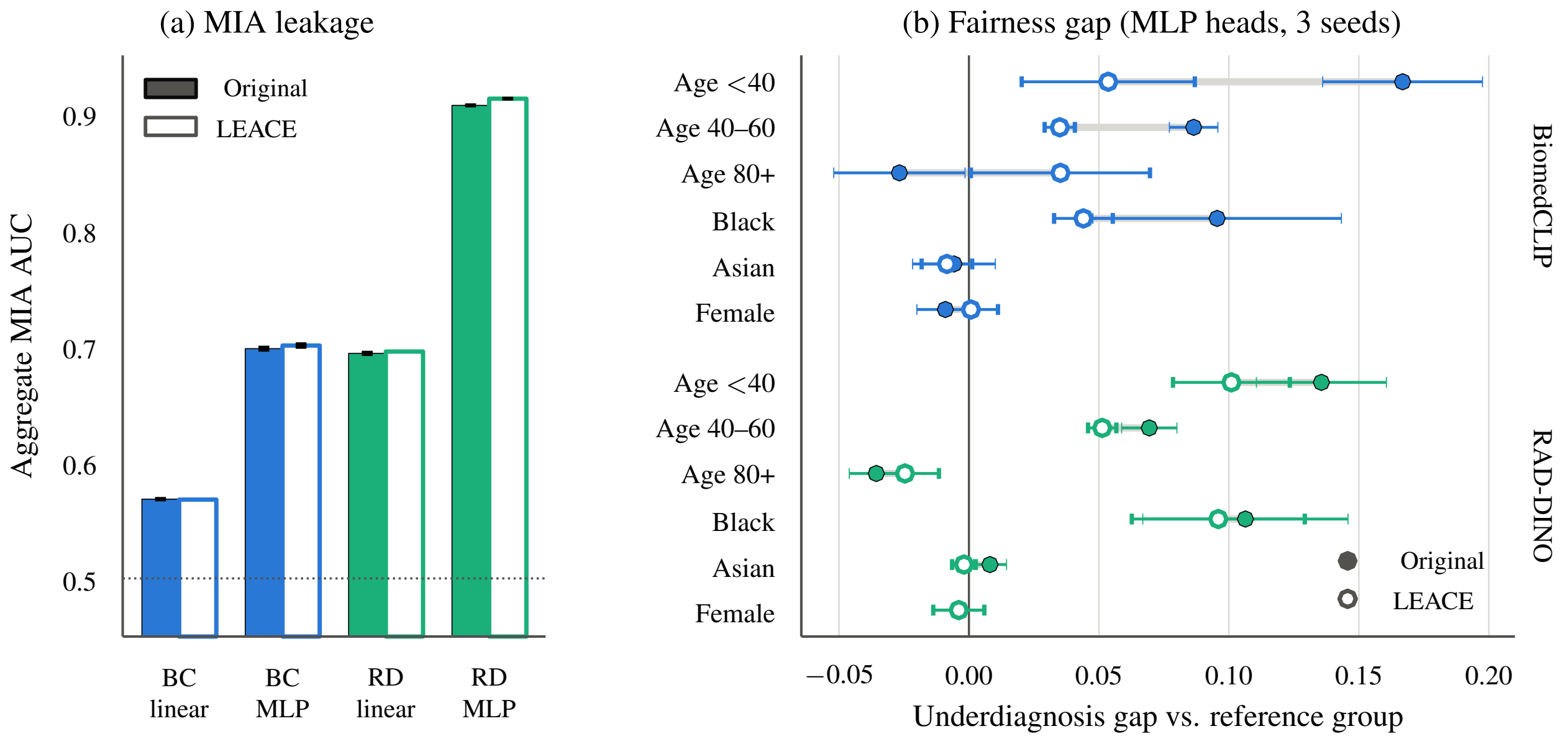}}
\caption{Effect of LEACE on leakage and fairness on CheXpert. (a) Aggregate MIA AUC before and after erasure (linear heads: one seed; MLP heads: mean over three seeds). (b) Underdiagnosis gap relative to the reference group for MLP heads. Values are means over three seeds, and bars show $\pm$1 s.d. across seeds.}\label{fig:leace}
\end{figure}

LEACE substantially reduced linearly decodable demographic information (Table~\ref{tab:leace}). Nonlinear probes still recovered sex, with an MLP-probe macro-AUROC of 0.994 for RAD-DINO, so the erasure was incomplete. Linear probes also retained some information (for example 0.683 for RAD-DINO sex). One possible reason is that LEACE was fitted on records, whereas the probes use patient-mean embeddings. Leakage did not change (Figure~\ref{fig:leace}a). The MIA AUC was 0.568 before and after erasure for the BiomedCLIP linear head. It moved from 0.698 to 0.700 for the BiomedCLIP MLP, from 0.694 to 0.695 for the RAD-DINO linear head and from 0.907 to 0.913 for the RAD-DINO MLP. The linear-head values come from one seed and the MLP values from three seeds. The cost in test AUROC was about 0.02 for BiomedCLIP and 0.01 for RAD-DINO.

Fairness, in contrast, responded to erasure (Figure~\ref{fig:leace}b). With MLP heads over three seeds, the underdiagnosis gap for patients under 40 years decreased from 0.167 to 0.054 on BiomedCLIP. The difference was 0.113 (95\% CI [0.073, 0.153]) and had the same direction in all three seeds. On RAD-DINO, the same gap decreased from 0.136 to 0.101 (difference 0.035, 95\% CI [0.027, 0.042]). The gap for patients aged 40 to 60 years decreased from 0.086 to 0.035 on BiomedCLIP and from 0.069 to 0.051 on RAD-DINO. For patients aged 80 years and over, the BiomedCLIP gap changed sign, from $-0.027$ to $+0.035$. The Black-White gap on BiomedCLIP shrank from 0.095 to 0.044, but its CI included zero, so we report this change as directional only. On RAD-DINO, this gap changed little, from 0.106 to 0.096. Sex and Asian gaps remained near zero throughout. These CIs are based on a $t$-interval over three seeds (two degrees of freedom) and are therefore crude.

Taken together, these results separate the two problems. Removing the age information that a linear model can read reduced age-related disparity, but it did not make patients less identifiable as training members.

\begin{table}[t]
\centering\small
\caption{Probe macro-AUROC on patient-mean embeddings before and after LEACE on CheXpert, using the probes described in Section~\ref{sec:sweeps}. The macro-AUROC chance level is 0.5.}\label{tab:leace}
\begin{tabular}{@{}llcccccc@{}}
\toprule
 & & \multicolumn{3}{c}{Linear probe} & \multicolumn{3}{c}{MLP probe}\\
\cmidrule(lr){3-5}\cmidrule(lr){6-8}
Encoder & Embedding & Age & Race & Sex & Age & Race & Sex\\
\midrule
BiomedCLIP & Original & 0.746 & 0.704 & 0.893 & 0.820 & 0.683 & 0.914\\
 & LEACE & 0.442 & 0.512 & 0.564 & 0.561 & 0.591 & 0.794\\
RAD-DINO & Original & 0.872 & 0.802 & 0.998 & 0.853 & 0.804 & 0.998\\
 & LEACE & 0.511 & 0.541 & 0.683 & 0.692 & 0.666 & 0.994\\
\bottomrule
\end{tabular}
\end{table}

\subsection{Results on NIH ChestX-ray14}\label{sec:nih}
NIH ChestX-ray14 reproduced the encoder and capacity ordering of leakage observed on CheXpert (Tables~\ref{tab:main} and~\ref{tab:regime}). The fairness results, however, differed. The gaps were small, and the reference age group was 40 to 60 years. The gap for patients aged 60 to 80 years ranged from $-0.057$ to $-0.098$, and the female gap ranged from $+0.008$ to $+0.041$. We therefore do not pool fairness findings across datasets.

%=====================================================================
\section{Discussion}\label{sec:discussion}
The intuitive view that encoding leads to exposure would justify a single remedy for both fairness and privacy, but our data do not support it. Encoding strength did not predict which patients leaked (Section~\ref{sec:equiv}), and erasing linearly decodable demographic information did not change leakage (Section~\ref{sec:leace}), although nonlinear demographic information remained. Instead, leakage depended on how the head was trained and on how atypical the patient was (Sections~\ref{sec:capacity} and~\ref{sec:who}). Erasure did, however, reduce age-related underdiagnosis. In our setting, encoding is therefore relevant to fairness without being relevant to membership exposure, and practitioners should audit the two risks separately.

The results have three practical implications for frozen-embedding pipelines. First, linear heads or shorter training reduced patient-level leakage substantially with little or no utility cost. For RAD-DINO on CheXpert, the MIA AUC fell from 0.907 for the 30-epoch MLP to 0.694 for the linear head with no loss of test AUROC, whereas BiomedCLIP on NIH ChestX-ray14 lost about 0.02 in test AUROC with a linear head or shorter training. Second, analyses based on the maximum over the records of a patient overstate the role of image count. Third, the most vulnerable patients were those with unusual findings or sparse embedding neighborhoods. These characteristics are not demographic attributes, which suggests that the patients most exposed to membership inference need not coincide with the groups at risk of underdiagnosis.

Our capacity result is consistent with Knolle et al.~\cite{knolle2026}, who found that leakage increases with model capacity and that the enrichment of groups in the patient tail partly tracks group size. We add two observations for frozen features. Training length mattered about as much as head width, and demographic encoding did not predict patient vulnerability once atypicality features were included in the model.

\subsection{Limitations}\label{sec:limitations}
Our conclusions apply to the setting we audited, which is small heads trained on frozen embeddings of chest radiographs from two public datasets. This setting corresponds to using a foundation model as a fixed feature extractor, and it allowed us to train 200-model ensembles with controlled membership. Whether the same dissociation holds for end-to-end fine-tuned models, other imaging modalities or differentially private training remains to be tested. Frozen embeddings can also discard small-lesion signal~\cite{muthyala2026}, so leakage at the backbone level may differ.

Several design choices define how far the intervention and encoding results extend. LEACE removes linear information only, and nonlinear probes still recovered some demographic signal (Table~\ref{tab:leace}). We fitted it on all records, including test patients, because we used it as an analytical intervention and not as a deployable mitigation. Our encoding scores are likewise linear and therefore measure linearly decodable demographic information. The capacity sweep and the linear-head LEACE runs used one seed, the confidence intervals of the fairness differences rest on three seeds, and the agreement between encoding and leakage at the encoder level rests on three encoders. We therefore treat these results as supporting evidence, and the main conclusions rest on the three-seed analyses. The 5-epoch setting is a fixed proxy for early stopping. The pretraining data of RAD-DINO include CheXpert, so its CheXpert results are an upper bound for a domain-matched encoder, and DINOv2 is only a partial control.

The demographic labels also bound the analysis. The CheXpert race labels are coarse, self-reported categories from the Stanford AIMI demographic file~\cite{gichoya2022} in which Hispanic ethnicity is folded into the race groups, and ``Unknown'' and ``Other'' were excluded. NIH ChestX-ray14 has no race labels, and its fairness pattern differs from that of CheXpert. We did not analyze intersectional subgroups.

%=====================================================================
\section{Conclusion}\label{sec:conclusion}
Demographic encoding, diagnostic fairness and membership vulnerability are separable in heads trained on chest X-ray foundation-model embeddings. Encoding strength was statistically equivalent to zero as a predictor of patient leakage in all 72 CheXpert tests. Linear erasure of demographics left leakage unchanged but reduced age-related underdiagnosis, although nonlinear demographic information remained. Leakage instead tracked head capacity, training length, embedding family and patient atypicality. Fairness and privacy should therefore be audited separately, and simple choices of head and training length can reduce exposure considerably. Future work should test whether these findings hold for end-to-end fine-tuned models, other imaging modalities and differentially private training.

\section*{Data and code availability}
{\small
Both datasets are publicly available. CheXpert~\cite{irvin2019} is released by the Stanford ML Group, and NIH ChestX-ray14~\cite{wang2017} is released by the NIH Clinical Center. Race labels for CheXpert were obtained from the demographic file released by Stanford AIMI \cite{gichoya2022}. We obtained copies through Kaggle: CheXpert-v1.0-small (\url{https://www.kaggle.com/datasets/ashery/chexpert}) and NIH Chest X-rays (\url{https://www.kaggle.com/datasets/nih-chest-xrays/data}). The source code is available at \url{https://github.com/SAMZ2610/privacy_bias_fairness}.
}

\section*{Declaration of generative AI in scientific writing}
During the preparation of this work, the authors used Claude Sonnet 5.5 to improve the English language, correct grammar, and enhance the overall flow of the text. After using this tool, the authors reviewed and edited the content as needed and take full responsibility for the final content of the publication.

\section*{Declaration of competing interest}
The authors declare no known competing financial interests or personal relationships that could have appeared to influence the work reported in this paper.

\section*{CRediT authorship contribution statement}
\textbf{Sahar Askari}: Conceptualization, Methodology, Software, Visualization, Investigation, Writing - Original Draft, Writing - Review \& Editing. \textbf{Amin Golnari}: Writing - Review \& Editing.

%Bibliography
\bibliographystyle{unsrt}  
\bibliography{references}  

\end{document}